\def\al{\alpha}

\def\be{\begin{equation}}
\def\ee{\end{equation}}
\def\ba{\begin{array}}
\def\ea{\end{array}}

\def\dps{\displaystyle}

\documentclass[groupedaddress,superscriptaddress,pra,showpacs,twocolumn,amsmath]{revtex4}
\usepackage{epsfig,graphicx} 
\usepackage{amsmath}
\usepackage[mathscr]{eucal}
\usepackage{amssymb}
\input amssym.def
\newtheorem{theorem}{Theorem}

\newtheorem{lemma}{Lemma}
\begin{document}

\title{Local Unitary Invariants of Generic Multi-qubit States}
\author{Naihuan Jing}
\email{jing@math.ncsu.edu}
\affiliation{School of Mathematics, South China University of Technology, Guangzhou, Guangdong 510640, China}
\affiliation{Department of Mathematics, North Carolina State University, Raleigh, NC 27695, USA}
\author{Shao-Ming Fei}
\affiliation{School of Mathematical Sciences, Capital Normal University, Beijing 100048, China}
\affiliation{Max-Planck-Institute for Mathematics in the Sciences, 04103 Leipzig, Germany}
\author{Ming Li}
\affiliation{Department of Mathematics, China University of Petroleum, Qingdao, Shandong 266555, China}
\author{Xianqing Li-Jost}
\affiliation{Max-Planck-Institute for Mathematics in the Sciences, 04103 Leipzig, Germany}
\author{Tinggui Zhang}
\affiliation{School of Mathematics and Statistics, Hainan Normal University, Haikou, Hainan 571158, China}
\begin{abstract}
We present a complete set of local unitary invariants for
generic multi-qubit systems which gives necessary and sufficient
conditions for two states being local unitary equivalent. These
invariants are canonical polynomial functions in terms of the generalized Bloch representation
of the quantum states. In particular, we prove that there are at most
12 polynomial local unitary invariants for two-qubit states
and at most 90 polynomials for three-qubit states. Comparison with
Makhlin's 18 local unitary invariants is given for two-quibit systems.
\end{abstract}

\maketitle

Local unitary equivalence is a foundational concept in quantum entanglement
and quantum information, as it provides the key symmetry
in classifying quantum entangled states of physical systems \cite{nielsen}.
Two quantum states are of the same nature in implementing quantum
information processing if they are equivalent under a
local unitary (LU) transformation, and many crucial properties such
as the degree of entanglement \cite{eof1,eof2}, maximal
violations of Bell inequalities \cite{bell1,bell2,bell3,bell4}, and
the teleportation fidelity \cite{tel1,tel2} remain invariant under
LU transformations. Moreover, quantum entanglement in multipartite qubits
has also figured prominently in many quantum
information processing such as one-way quantum computing, quantum error
correction and quantum secret sharing \cite{1,2,3,4}. For this reason, it has been a key
problem to find a complete and operational procedure to distinguish two quantum states under LU transformations.

In \cite{makhlin}, Makhlin presented a complete set of 18 polynomial LU invariants
for classifying two-qubit states. There are numerous
results on LU invariants for three qubits states \cite{linden}, some general
mixed states \cite{L1, SFG,
SFW, SFY, zhou}, tripartite pure and mixed states
\cite{SCFW}. A theoretical method to reduce the problem to pure $n$-qubit states was
proposed in \cite{mqubit}, and later generalized to arbitrary dimensions in \cite{bliu}.
From a different viewpoint, \cite{Li2} gave
a procedure to find the LU operator for multi-qubits using the core tensor
method.
Very recently a method to judge LU equivalence for multi-qubits \cite{Ma} was also
proposed and more generally SLOCC invariants
for multi-partite states are found \cite{JLLZF}. Nevertheless, it remains a wild problem
to find a complete set of invariants to
answer the LU question except for two qubit cases.
Even for two partite cases it is also desirable to find an alternative
set of invariants to judge LU equivalence, as the original Makhlin invariants contain some nontrivial tensor
vectors.

In this article, we propose a brand new method to quantify polynomial LU invariants
for multi-qubit systems in an operational way. For the special case of two-qubit systems, our method is more efficient
and needs fewer invariants than that in \cite{makhlin} in general. In fact, we
show that many invariants in \cite{makhlin} are consequences of other
invariants, and there are at most 12 invariants to
determine the LU equivalence for two-qubit states. We prove for the first
time that there are at most 90 invariants for generic mixed 3-qubit states. We also
propose an operational method to derive a list of polynomial invariants for generic multi-qubit states.
We remark that the invariants can not be derived from \cite{Li2} as the latter
aimed to compute the LU operator for two equivalent multi-qubits, while our current work
takes a different strategy to seek a complete set of polynomial invariants.

We start our discussion to
express an $N$-qubit state $\rho$ in terms of Pauli matrices $\sigma_\alpha$, $\alpha=1, 2, 3$,
\be\label{rhoN}
\ba{rcl}
\rho&=&\displaystyle\frac{1}{2^N} I^{\otimes N}
+\sum_{j_{1}=1}^N\sum_{\alpha_{1}=1}^3 T^{\alpha_{1}}_{j_{1}}\,\sigma_{\alpha_{1}}^{j_{1}}\\[4mm]
&&+\displaystyle\sum_{1\leq j_{1}<j_{2}\leq N}\,\sum_{\alpha_{1},\alpha_{2}=1}^3T^{\alpha_{1}\alpha_{2}}_{j_{1}j_{2}}
\,\sigma_{\alpha_{1}}^{j_{1}}\sigma_{\alpha_{2}}^{j_{2}}+\cdots \\[4mm]
&&\displaystyle +\sum_{1\leq j_{1}<\cdots<j_{M}\leq N}\sum_{\alpha_{1},\cdots,\alpha_{M}=1}^3
T^{\alpha_{1}\alpha_{2}\cdots\alpha_{M}}_{j_{1}j_{2}\cdots j_{M}}\,\sigma_{\alpha_{1}}
^{j_{1}}\cdots\sigma_{\alpha_{M}}^{j_{M}}\\[4mm]
&&+\cdots +\dps\sum_{\alpha_{1},\alpha_{2},\cdots,\alpha_{N}=1}^3
T^{\alpha_{1}\alpha_{2}\cdots\alpha_{N}}_{12\cdots N}\,\sigma_{\alpha_{1}}^{1}\sigma_{\alpha_{2}}^{2}\cdots\sigma_{\alpha_{N}}^{N},
\ea
\ee
where $I$ is the $2\times 2$ identity matrix, $\sigma_{\alpha_{k}}^{j_{k}}=I\otimes
I\otimes\cdots\otimes \sigma_{\alpha_{k}}\otimes
I\otimes\cdots\otimes I$ with $\sigma_{\alpha_{k}}$ at the $j_k$-th position and
\begin{eqnarray}\label{T}
T^{\alpha_{1}\alpha_{2}\cdots\alpha_{M}}_{j_{1}j_{2}\cdots j_{M}}=\frac1{2^N}{\rm Tr}[\rho\,\sigma_{\alpha_{1}}
^{j_{1}}\sigma_{\alpha_{2}}^{j_{2}}\cdots\sigma_{\alpha_{M}}^{j_{M}}],~~~~M\leq N,
\end{eqnarray}
are real coefficients. In particular, $T_{j}=(T^{1}_{j},T^2_{j},T^3_{j})$, $j=1,...,N$, are three dimensional vectors,
$T_{jk}=(T^{\alpha_{1}\alpha_{2}}_{jk})$, $1\leq j<k\leq N$, are $3\times 3$ matrices.
Generally, $T_{j_{1}j_{2}\cdots j_{M}}=(T^{\alpha_{1}\alpha_{2}\cdots\alpha_{M}}_{j_{1}j_{2}\cdots j_{M}})$ are tensors.


Let $\rho$ and $\rho'$ be two $N$-qubit mixed states. They are called local unitary equivalent if
\be\label{lueq}
\rho'=(U_1\otimes...\otimes U_N)\rho (U_1\otimes...\otimes U_N)^\dag
\ee
for some unitary operators $U_i\in SU(2)$, $i=1,2,...,N$, where $\dag$ denotes transpose and conjugate.

\begin{lemma} Two mixed states $\rho$ and $\rho'$ are local unitary
equivalent if and only if there are special
orthogonal matrices $O_1, \cdots, O_N\in SO(3)$ such that
\begin{equation}\label{LU1}
(O_{j_1}\otimes\cdots \otimes O_{j_k})T_{j_1\cdots j_k}
=T'_{j_1\cdots j_k}
\end{equation}
for any $1\leq j_1< \cdots <j_k\leq N$, $k=1,2,...,N$.
\end{lemma}

\noindent{\bf Proof}. The group $SU(2)$ acts on the real vector space
spanned by $\sigma_i$, $i=1, 2, 3$ via \cite{JLLZF}:
\be\label{jllzf}
U_i\sigma_kU_i^{\dagger}=\sum_{j=1}^3O_{kl}\sigma_l,
\ee
where $O=(O_{kl})$ belongs to $SO(3)$. From (\ref{rhoN}), (\ref{lueq}) and (\ref{jllzf}) one gets the tensor
relation (\ref{LU1}). Note that this action realizes the well-known double-covering map $SU(2)\rightarrow SO(3)$.
The sufficiency then follows from the fact that $SU(2)$ is the universal double covering of $SO(3)$. \hfill
\rule{1ex}{1ex}

{\it Two-qubit states:}~~
To derive explicitly the invariants under the transformation (\ref{lueq}),
we first consider the two-qubit case. From (\ref{rhoN}) a two-qubit state is given by
the $3$-dimensional real column vectors $T_1$, $T_2$, and the real $3\times 3$-matrix $T_{12}$.
Two states $\rho$ and $\rho'$ are local unitary equivalent if and if there are $SO(3)$ operators $O_1$ and $O_2$
such that
\be\label{LU2}
\ba{rcl}
T'_1&=&O_1T_1, \quad T'_2=O_2T_2,\\[2mm]
T'_{12}&=&(O_1\otimes O_2)T_{12}=O_1T_{12}O_2^t,
\ea
\ee
where $t$ denotes the transpose of a matrix.

We introduce the following sets of $3$-dimensional real column vectors:
\begin{align*}
\langle\mathcal O_1\rangle&=\{T_1, T_{12}T_2, T_{12}T_{12}^tT_1, T_{12}T_{12}^t
T_{12}T_2,\cdots\}\subset\mathbf R^3,\\
\langle\mathcal O_2\rangle&=\{T_2, T_{12}^tT_1, T_{12}^tT_{12}T_2, T_{12}^tT_{12}
T_{12}^tT_1,\cdots\}\subset\mathbf R^3,
\end{align*}
which are respectively generated by the $(T_{12}T_{12}^t)$-orbit of $\{T_1, T_{12}T_2\}$
and the $(T_{12}^tT_{12})$-orbit of $\{T_2, T_{12}^tT_1\}$. Here
$(g)$ denotes the cyclic group generated by $g$.
By the Cayley-Hamilton theorem the minimal polynomials of
$T_{12}T_{12}^t$ and $T_{12}^tT_{12}$ have degree $\leq 3$, therefore
it is enough to use elements in the orbits up to the quadratic powers.
It is straightforward to verify that all the vectors in
$\langle\mathcal O_1\rangle$ are transformed to $O_1 \langle\mathcal O_1\rangle$
under the transformation (\ref{LU2}),
while all the vectors in
$\langle\mathcal O_2\rangle$ are transformed into $O_2 \langle\mathcal O_2\rangle$.
Moreover, there are at most three linear independent vectors in
$\langle\mathcal O_i\rangle$, $dim\langle O_i\rangle\leq 3$, $i=1,2$.
We say that a two-qubit state is {\it generic} if $dim\langle O_1\rangle=dim\langle O_2\rangle=3$.
For simplicity, we only deal with generic cases in the following.
The non-generic (degenerate) cases can be studied in details too, see remarks after
the proof of Theorem 1.

Let $\{\mu_1, \cdots, \mu_6\}$ and $\{\nu_1, \cdots, \nu_6\}$ denote the first six
(spanning) vectors in $\langle\mathcal O_1\rangle$ and $\langle\mathcal O_2\rangle$, respectively.
We first give a general result using all the spanning vectors.

\begin{theorem} \label{T:bipartite}  Two generic two-qubit states are
local unitary equivalent if and only if they have the same values of the following
invariant polynomials:
\be\label{ip}
\ba{l}
\langle \mu_i, \mu_j\rangle ,\quad \langle \nu_i, \nu_j\rangle,~~~i\leq j=1,2, \cdots, 6. \\[2mm]
tr(T_{12}T_{12}^t)^{\alpha},~~~ \alpha=1,2,3.
\ea
\ee
\end{theorem}

\noindent{\bf Proof}. By using the relations in (\ref{LU2}), it is direct to verify that the quantities given in (\ref{ip}) are invariants under
local unitary transformations.

For generic states, the matrix $T_{12}T_{12}^t$ is nonsingular, so is $T^t_{12}T_{12}$ by
trace property. Thus $T_{12}$ and $T_{12}^t$ are also nonsingular. We notice that
$\langle\mathcal O_1\rangle \overset{\tiny T_{12}^t}{\longrightarrow}\langle\mathcal O_2\rangle$
and $\langle\mathcal O_2\rangle\overset{\tiny T_{12}}{\longrightarrow} \langle\mathcal O_1\rangle$
as subsets or subspaces, therefore $\langle\mathcal O_1\rangle\simeq\langle\mathcal O_2\rangle=\mathbf R^3$
for generic states, and a basis of $\mathbf R^3$ can be pared down from the vectors of $\langle\mathcal O_1\rangle$
or $\langle\mathcal O_2\rangle$ by assumption.

Assuming that two generic two-qubit states $\rho$ and $\rho'$ have the same values of the
invariant polynomials (\ref{ip}), namely, the inner products of any two vectors
in $\langle\mathcal O_i\rangle$ are invariant under $\rho\to\rho'$, one has that
there must exist an orthogonal matrix $O_i$ such that
\begin{equation*}
O_i\langle\mathcal O_i\rangle=\langle\mathcal O_i'\rangle.
\end{equation*}
In particular $O_iT_i=T'_i$. Then we can build the following commutative diagram:
$$
\begin{array}[c]{rlr}
\langle\mathcal O_1\rangle
&~\stackrel{O_1}{\longrightarrow}~& \langle\mathcal O_1'\rangle\\[2mm]
\left\downarrow\rule{0cm}{0.5cm}\right.\scriptstyle{T_{12}^t}&&\left\downarrow\rule{0cm}{0.5cm}\right.\scriptstyle{T_{12}^{'t}}\\[2mm]
\langle\mathcal O_2\rangle &~\stackrel{O_2}{\longrightarrow}~& \langle\mathcal O_2'\rangle
\end{array}
$$
Consequently $T_{12}^{'t}O_1=O_2T_{12}^t$ in $End(\mathbf R^3)$, or $T_{12}'=O_1T_{12}O_2^t$.
Therefore, $\rho$ and $\rho'$ are local unitary equivalent. \hfill\rule{1ex}{1ex}

{\it Remark}. In the above discussions we are only concerned with the generic case. For degenerate cases,
one needs to analyze case by case. For instance, let us consider the case $T_1=T_2=0$.
In this case, $dim\langle O_i\rangle=0$, $i=1,2$. The only invariants left are $tr(T_{12}T_{12}^t)^{\alpha}$, $\alpha=1,2,3$.
Note that
\begin{equation}\label{e:ps}
p_{\alpha}=tr(T_{12}T_{12}^t)^{\alpha}=\sum_{i=1}^3\lambda_i^{\alpha}
\end{equation}
is the $\al$th-power sum
of the eigenvalues of $T_{12}T_{12}^t$. A well-known result of symmetric polynomials implies
that any $p_{\alpha}$ $(\alpha\geq 4)$ is an algebraic function of $p_1, p_2$, and $p_3$. For example,
$p_4=\frac16 p_1^4-p_1^2p_2+\frac12 p_2^2+\frac43 p_1p_3$.
Hence $tr(T_{12}T_{12}^t)^{\alpha}$ are invariants for any $\alpha\geq 1$.
By \cite{Sp} if two states $\rho$ and $\rho'$ have the same values of $tr(T_{12}T_{12}^t)^{\alpha}$,
there exists a unitary matrix $U$ such that $T'_{12}T_{12}^{'t}=UT_{12}T_{12}^{t}U^{\dagger}$, which means that $T_{12}T_{12}^{t}$ and $T'_{12}T_{12}^{'t}$
have identical eigenvalues. Both $T_{12}T_{12}^{t}$ and $T_{12}'T_{12}^{'t}$ are similar to $diag(\lambda_1, \lambda_2, \lambda_3)$. Then
there exists an $O_1\in SO(3)$ such that
$T'_{12}T^{'t}_{12}=O_1 T_{12}T_{12}^tO_1^t$. Similarly there exists $O_2$ such that $T^{'t}_{12}T'_{12}=O_2 T_{12}^tT_{12}O_2^t$.
Subsequently $T'_{12}=O_1T_{12}O_2^t$ for some $O_1$ and $O_2$, so $\rho$ and $\rho'$ are local unitary equivalent.

We now sharpen the result of Theorem \ref{T:bipartite}. Since there are at most three linearly
independent 3-dimensional vectors of $\mu_i$ and $\nu_i$ in (\ref{ip}) respectively,
one can apply Theorem \ref{T:bipartite} to the basis vectors.
The standard Gaussian elimination
on the matrix $[\mu_1, \cdots, \mu_6]$ can pare down the column vectors into a basis $\{\mu_{i_1}, \mu_{i_2}, \mu_{i_3}\}$ of $\langle O_1\rangle$, where
$\{i_1, i_2, i_3\}\subset \{1, 2, \cdots, 6\}$.
This means that
the number of independent invariants that are used to judge the local unitary equivalence
of two generic two-qubit states
is at most 15 in general (instead of 33 as is Theorem \ref{T:bipartite}). In fact, further analysis can reduce the
number to at most 12 polynomial invariants.

\begin{theorem} \label{T:bipartite2} Two generic two-qubit states
are local unitary equivalent if and only if they have the same values for the following 12 invariants:
\begin{align}\label{ip2}
&\langle T_1, (T_{12}T_{12}^t)^\beta T_1\rangle,~~~ \ \langle T_2, (T_{12}^tT_{12})^\beta T_2\rangle,\\ \label{ip2b}
&\langle T_1, (T_{12}T_{12}^t)^\beta \,T_{12}T_2\rangle, ~~~ \beta =0, 1, 2,\\ \label{ip2c}
&tr(T_{12}T_{12}^t)^{\alpha},~~~ \alpha=1,2,3.
\end{align}
\end{theorem}

\noindent{\bf Proof}. The set $\langle\mathcal O_1\rangle$ is a
union of two orbits $(T_{12}T_{12}^t)\cdot T_1$ and
$(T_{12}T_{12}^t)\cdot T_{12}T_2$. The independent inner
products given in Theorem \ref{T:bipartite} are $\langle T_1, (T_{12}^tT_{12})^\beta T_1\rangle$,
$\langle T_1, (T_{12}T_{12}^t)^\beta T_{12}T_2\rangle$ for $\beta=0,1,2,3$,
due to the Cayley-Hamilton theorem and the fact that
$\langle T_{12}u, v\rangle=\langle u, T_{12}^tv\rangle$
for any vectors $u, v$ ($T_{12}$ ia a real matrix).
Similarly the orbit $\langle\mathcal O_2\rangle$ will only
contribute the remaining independent inner products $\langle T_2, (T_{12}^tT_{12})^\beta T_2\rangle$,
$\beta=0, 1, 2, 3$.

We claim that the 3 invariants with $\beta=3$ are not needed if the traces (\ref{ip2c}) are
known.
The Cayley-Hamilton theorem says that
\begin{align}\label{e:CH}
(T_{12}T_{12}^t)^3=e_1(T_{12}T_{12}^t)^2-e_2(T_{12}T_{12}^t)+e_3I,
\end{align}
where $e_i$ are the elementary symmetric polynomials in the eigenvalues $\lambda_i$.
By the fundamental theorem of symmetric polynomials,
the $e_i$ can be expressed as classical polynomials in the traces $p_{\alpha}$ (\ref{e:ps}),
i.e. $e_i$ are classical invariant polynomials of the density matrix:
\begin{align}\label{e:classical1}
e_1&=p_1, \quad
e_2=\frac12(p_1^2-p_2),\\ \label{e:classical2}
e_3&=\frac16(p_1^3-3p_2p_1+2p_3).
\end{align}
Plugging (\ref{e:CH}) into the three invariants
$\langle T_1, (T_{12}^tT_{12})^3 T_1\rangle$ etc.,
we see that they are given by linear combinations of the invariants
(\ref{ip2}-\ref{ip2b}) with fixed coefficients of the classical invariant
polynomials (\ref{e:classical1}-\ref{e:classical2}) of the density matrix, so they are redundant.
\hfill\rule{1ex}{1ex}

As we commented above if we use the Gaussian elimination
we also worry about just 12 invariants. i.e., if
we add $\beta=3$ in the first set of invariants (\ref{ip2}-\ref{ip2b})
for the 3 basis vectors we can waive the trace identities.
Hence the total number of invariants is at most $12$ either way.
We still include the trace identities (\ref{ip2c}) for the sake of general (non-generic) cases.
\smallskip

{\it Multi-qubit case:} To simplify presentation, we introduce the following
notation: $T_{ij}=T_{ji}^t$. We say that
a word of $T_i, T_{ij}$ is admissible if the
adjacent subindices match. For example, $T_{12}T_2, T_{12}T_{21}T_1T_{12}$ are admissible ones.

We first consider the three-qubit case to present our general results.
In this case, corresponding to (\ref{rhoN}), a quantum state has the form:
\begin{equation}
\rho=\frac18I+\sum_{i=1}^3T_i\,\sigma^{(i)}+\sum_{i<j}^3T_{ij}\,\sigma^{(i)}\sigma^{(j)}
+T_{123}\,\sigma^{(1)}\sigma^{(2)}\sigma^{(3)}.
\end{equation}
If two states $\rho$ and $\rho'$ are local unitary equivalent, then there
are orthogonal matrices $O_i\in SO(3)$ such that
\begin{align}\label{LU3a}
&T'_1=O_1T_1, \quad T'_2=O_2T_2, \quad T'_3=O_3T_3,\\ \label{LU3b}
&T'_{12}=O_1T_{12}O_2^t,~ T'_{13}=O_1T_{12}O_3^t,~ T'_{23}=O_2T_{23}O_3^t,\\ \label{LU3c}
&T'_{123}=(O_1\otimes O_2\otimes O_3)T_{123}.
\end{align}
It is known \cite{L1} that the last relation (\ref{LU3c}) is equivalent to
either of the following two relations:
\begin{equation}\label{LU3d}
T'_{123}=O_1T_{123}(O_2\otimes O_3)^t,~ T'_{123}=(O_1\otimes O_2)T_{123}O_3^t.
\end{equation}
Here $T_{123}$ is understood as the bipartition $T_{1|23}$ (resp. $T_{12|3}$) in the first (resp. 2nd) equation of (\ref{LU3d}).
To state our results we introduce two subsets of vectors:
\begin{widetext}
\begin{align*}
\langle\mathcal O_1\rangle_{1|23}&=\{T_1, T_{123}(T_{23}T_{23}^t)^\beta T_{23}, T_{123}T_{123}^tT_1, T_{123}T_{123}^t
T_{123}(T_{23}T_{23}^t)^\beta T_{23}, (T_{123}T_{123}^t)^2T_1, \cdots\}\subset\mathbf R^3,\\
\langle\mathcal O_2\otimes \mathcal O_3\rangle_{1|23}&=
\{T_{23}, T_{123}^tT_1, T_{123}^tT_{123}(T_{23}T_{23}^t)^\beta T_{23}, T_{123}^tT_{123}
T_{123}^tT_1, (T_{23}T_{23}^t)^\beta T_{23},\cdots\}\subset \mathbf R^9\simeq\mathbf R^3\otimes \mathbf R^3,
\end{align*}
\end{widetext}
where $\beta=0, \cdots, 3$, which are respectively the $(T_{123}T_{123}^t)$-orbit of $\{T_1, T_{123}(T_{23}T_{23}^t)^\beta T_{23}|\,\beta=0,1,2,3\}$
and the $(T_{123}^tT_{123})$-orbit of $\{(T_{23}T_{23}^t)^\beta T_{23}, T_{123}^tT_1|\,\beta=0,1,2,3\}$. Here
$T_{23}$ is taken as its (column) vector realignment in $\mathbf R^9$ and $T_{123}$
is folded as a $3\times 9$-matrix, by viewing $T_{123}$ as the bipartition $1|23$
and $T_{123}^t$ is the transpose with respect to such partition. As before
we also use the same symbols for the corresponding real subspaces.

Similarly, by permuting the indices
we define $\langle \mathcal O_2\rangle:=\langle\mathcal O_2\rangle_{2|31}$ and $\langle \mathcal O_3\rangle
:=\langle\mathcal O_3\rangle_{3|12}$ to be
the
$(T_{231}T_{231}^t)$-orbit of $\{T_2, T_{231}(T_{31}T_{31}^t)^\beta T_{31}|\,\beta=0,1,2,3\}$
and the $(T_{123}^tT_{123})$-orbit of $\{T_{3}, T_{312}(T_{12}T_{12}^t)^\beta T_{12}|\,\beta=0,1,2,3\}$ respectively.
Here the $3\times 9$-matrix $T_{231}$ (resp. $T_{312}$) is the realignment of $T_{123}$ with respect to the partition
of $\{123\}$ into $\{2|31\}$ (resp. $\{3|12\}$).
Let $\langle\mathcal O_1\rangle=\{\mu_1, \mu_2, \mu_3\}$,
$\langle\mathcal O_2\rangle=\{\nu_1, \nu_2, \nu_3\}$
and $\langle\mathcal O_3\rangle=\{\lambda_1, \lambda_2, \lambda_3\}$;
$\langle\mathcal O_2\otimes \mathcal O_3\rangle_{1|23}=\{\alpha_1, \alpha_2, \ldots, \alpha_9\}$,
$\langle\mathcal O_3\otimes \mathcal O_1\rangle_{2|31}=\{\beta_1, \beta_2, \ldots, \beta_9\}$,
and $\langle\mathcal O_1\otimes \mathcal O_2\rangle_{3|12}=\{\gamma_1, \gamma_2, \ldots, \gamma_9\}$.

\begin{theorem}\label{T:tripartite}  A three-qubit state
$\rho$ is local unitary equivalent to a three-qubit state $\rho'$ if and only if the respective
invariant polynomials are equal:
\begin{align}\nonumber
\langle \mu_i, \mu_j\rangle &=\langle \mu'_i, \mu'_j\rangle,~~~
\langle \nu_i, \nu_j\rangle=\langle \nu'_i, \nu'_j\rangle, \\\nonumber
\langle \lambda_i, \lambda_j\rangle&=\langle \lambda'_i, \lambda'_j\rangle,~~~ 1\leq i\leq j\leq 3\\
\langle \alpha_k, \alpha_l\rangle &=\langle \alpha'_k, \alpha'_l\rangle,~~~ 1\leq k\leq l\leq 9\\\nonumber
\langle \beta_k, \beta_l\rangle &=\langle \beta'_k, \beta'_l\rangle,~~~ 1\leq k\leq l\leq 9\\\nonumber
\langle \gamma_k, \gamma_l\rangle &=\langle \gamma'_k, \gamma'_l\rangle,~~~ 1\leq k\leq l\leq 9\\\nonumber
\end{align}
\end{theorem}
\noindent{\bf Proof}. By the result of two-qubit case,
the invariance of inner products
of vectors in $\langle\mathcal O_i\rangle$
implies the existence of orthogonal matrices
$O_i$, $i=1,2,3$ such that
Eqs. (\ref{LU3a}-\ref{LU3b}) hold. Thus we are left
to show that the orthogonal matrices
$O_i$ also satisfy
Eq. (\ref{LU3c}) or equivalently Eq. (\ref{LU3d}).

We use a similar method of Theorem \ref{T:bipartite} to show this by
viewing the three-qubit state $\rho$ as
a bi-partite one on $\mathbf C^3\otimes \mathbf C^9$
and partition the hyper-matrix $T_{123}$ as a
rectangular matrix $T_{1|23}$.
Then the $3\times 9$-matrix $T_{123}$
maps the subset $\langle\mathcal O_2\otimes \mathcal O_3\rangle_{1|23}$
into the subset $\langle\mathcal O_1\rangle_{1|23}$ by
left multiplication.

We have already seen that
there exists an orthogonal matrix $O_i$ such that
\begin{equation*}
O_i\langle\mathcal O_i\rangle=\langle\mathcal O_i'\rangle.
\end{equation*}
and Eqs. (\ref{LU3b}) hold. Then we can directly verify that the following
diagram is commutative:
$$
\begin{array}[c]{ccc}
\langle\mathcal O_2\otimes\mathcal O_3\rangle
&\stackrel{O_2\otimes O_3}{\longrightarrow}& \langle\mathcal O'_2\otimes\mathcal O'_3\rangle\\[2mm]
\left\downarrow\rule{0cm}{0.5cm}\right.\scriptstyle{T_{123}}&&\left\downarrow\rule{0cm}{0.5cm}\right.
\scriptstyle{T_{123}^{'}}\\[2mm]
\langle\mathcal O_1\rangle_{1|23}
&\stackrel{O_1}{\longrightarrow}& \langle\mathcal O_1'\rangle_{1|23}\
\end{array}
$$
Consequently $T_{123}^{'}(O_2\otimes O_3)=O_1T_{123}$ in $End(\mathbf R^3\otimes \mathbf R^3)$, or $T_{123}'=O_1T_{123}(O_2\otimes O_3)^t$.
\hfill\rule{1ex}{1ex}

The following result shows that there are at most 90 invariants to judge LU equivalence
for two three-qubit states.

\begin{theorem} \label{T:tripartite2} Two generic three-qubit states
are local unitary equivalent if and only if they have the same values of the following invariants:
\begin{align}\nonumber
&\langle T_1, (T_{12}T_{12}^t)^\alpha T_1\rangle, ~~~ \langle T_2, (T_{12}^tT_{12})^\alpha T_2\rangle,\\\nonumber
&\langle T_1, (T_{12}T_{12}^t)^\alpha \,T_{12}T_2\rangle, \\\nonumber
&tr(T_{12}T_{12}^t)^\beta, ~~ tr(T_{13}T_{13}^t)^\beta, ~~ tr(T_{23}T_{23}^t)^\beta, \\
&\langle T_1, (T_{1|23}T_{1|23}^t)^k T_1\rangle, ~~ \langle T_2, (T_{2|31}T_{2|31}^t)^k T_2\rangle,\\\nonumber
&\langle T_3, (T_{3|12}T_{3|12}^t)^k T_3\rangle; ~~~  \langle T_{23}, (T_{1|23}^tT_{1|23})^k T_{23}\rangle,\\\nonumber
&\langle T_{23}, (T_{1|23}^tT_{123})^k T_{1|23}^tT_1\rangle, \label{tri3}\\\nonumber
&tr(T_{1|23}^tT_{1|23})^l, ~~ tr(T_{2|31}^tT_{2|31})^l, ~~ tr(T_{3|12}^tT_{3|12})^l.
\end{align}
where $\alpha=0,1,2$; $\beta=1, 2, 3$ and $k=0, 1, \cdots, 8$; $l=1, \cdots, 9$.
\end{theorem}

The above criteria can be generalized to multi-qubits.
Define $\langle\mathcal O_i\rangle=\langle T_i, T_{n\cdots n1\cdots i-1|i}T_n, \cdots\rangle\subset \mathbf R^3$
as the $(T_{\hat ii}^tT_{\hat ii})$-orbit, where $\hat i=1\cdots\hat{i}\cdots n$ means the index $i$ is absent.
In general for any strict sequence $\mathbf i=(i_1\cdots i_k)$ (i.e. distinct $i_j$'s), we define
the $(T_{\hat{\mathbf i}\mathbf i}^tT_{\hat{\mathbf i}\mathbf i})$-orbit
$\langle O_{\mathbf i}\rangle=\langle T_{\mathbf i}, \cdots\rangle$,
where the admissible generating words have only $ i$ when crossing out redundant strings.
e.g., $T_{312}T_{12}T_1$ is a word of indices $3,1$ when crossing out $12$. Then we have the following
result. Let $\langle\mathcal O_1\rangle=\{\mu_1, \mu_2, \ldots, \mu_m\}$,
$\langle\mathcal O_2\rangle=\{\nu_1, \nu_2, \ldots, \nu_m\}, \cdots$,
$\langle\mathcal O_N\rangle=\{\lambda_1, \lambda_2, \ldots, \lambda_m\}$,
and more generally, for any strict sequence $\mathbf i$, let $\langle O_{\mathbf i}\rangle=\{\tau_1, \cdots, \tau_n\}$, where $n=n(\mathbf i)$. Let's list these $\langle O_{\mathbf i}\rangle$ as
$\langle O_{\mathbf i_j}\rangle=\{\tau^{(j)}_1, \cdots, \tau^{(j)}_{m_j}\}$, $j=1, \cdots, M$.

\begin{theorem}\label{T:multipartite}
Two generic multi-qubit states $\rho$ and $\rho'$ are local unitary equivalent if and only if the respective
invariant polynomials are equal:
\begin{align}\nonumber
\langle \tau_{i}^{(1)}, \tau_{j}^{(1)}\rangle&=\langle \tau_{i}^{(1)'}, \tau_{j}^{(1)'}\rangle, \cdots\\
\langle \tau_{i}^{(M)}, \tau_{j}^{(M)}\rangle &=\langle \tau_{i}^{(M)'}, \tau_{j}^{(M)'}\rangle,
\end{align}
where each pair of indices $(i, j)$ are such that
$1\leq i, j\leq m(\mathbf i)$ for the sequences $\mathbf i_1, \cdots, \mathbf i_M$.
\end{theorem}
\noindent{\bf Proof}. We use induction on $n$ to reduce the problem to $(n-1)$-partite qubits.
Note that for any sequence $\mathbf i$ of indices for $n$-partite state, we can
view the elements in $\langle O_{\mathbf i}\rangle$ as $\langle O_{\mathbf i'}\otimes O_j\rangle$
where $\mathbf i'$ is obtained by realignment of the Block matrix with respect to the index $j$,
and $\mathbf i'$ is obtained from $\mathbf i$ after the realignment. Then we can use the similar
commutative diagram
$$
\begin{array}[c]{ccc}
\langle\mathcal O_{2}\otimes\mathcal O_{3\cdots n}\rangle
&\stackrel{O_{2}\otimes O_{3\cdots n}}{\longrightarrow}& \langle\mathcal O'_{2}\otimes\mathcal O'_{3\cdots n}\rangle\\[2mm]
\left\downarrow\rule{0cm}{0.5cm}\right.\scriptstyle{T_{1\cdots n}}&&\left\downarrow\rule{0cm}{0.5cm}\right.
\scriptstyle{T_{1\cdots n}^{'}}\\[2mm]
\langle\mathcal O_1\rangle_{1|2\cdots n}
&\stackrel{O_1}{\longrightarrow}& \langle\mathcal O_1'\rangle_{1|2\cdots n}\
\end{array}
$$
to get $T_{12\cdots n}^{'}(O_2\otimes O_{3\cdots n})=O_1T_{12\cdots n}$ in $End(\mathbf R^3\otimes \mathbf R^{3(n-2)})$, or $T_{12\cdots n}'=O_1T_{12\cdots n}(O_2\otimes O_{3\cdots n})^t$. Here
$O_{3\cdots n}$ is an orthogonal matrix in the bigger orthogonal group.
Then we use the induction to argue further for the matrix $T_{1|2\cdots n}$ viewed as a reduced matrix for $(n-1)$-partite state to get
the final result.
\hfill\rule{1ex}{1ex}

{\it Conclusions and Remarks:} It is a basic and fundamental
question to classify quantum states under local unitary operations.
The problem has been figured out in \cite{mqubit,bliu} for pure multipartite quantum states.
However, it is much more difficult to
classify mixed quantum states under LU transformations.
Operational methods have been presented only for
non-degenerate bipartite states. Although the authors in \cite{jpa}
have shown that the problem of mixed states can be reduced to
one of pure states in terms of the purification of mixed states
mathematically, the protocol is far from being operational.
We have provided an operational way to verify and classify
quantum states by using the generalized Bloch representation in
terms of the generators of $SU(2)$. We remark that \cite{Li2} gives a practical procedure
to compute the LU operator for two equivalent multi-qubits, but it
can not derive the polynomial invariants from the procedure, as it is
based on a different strategy. In our current approach we set our goal to
write down a set of simple invariants with which two states can be easily checked
if they are LU equivalent. Since the coefficients (tensors) in the
representation can be determined directly by measuring some local quantum
mechanical observables-Pauli operators, the method is experimentally feasible. Our
criterion is both sufficient and necessary for generic multi-qubit quantum
systems, thus gives rise to a complete classification of multi-qubit
generic quantum states under LU transformations.

\smallskip
\noindent{\bf Acknowledgments}\, \, The work is supported by
NSFC (11105226, 11271138, 11275131), CSC, Simons Foundation 198129, Humboldt
Foundation, the Fundamental Research Funds for
the Central Universities (12CX04079A, 24720122013), and Research
Award Fund for outstanding young scientists of Shandong Province
(BS2012DX045).

\end{document}